\documentclass[conference]{IEEEtran}
\usepackage{amsmath,amssymb,amsfonts,mathrsfs,bm}
\usepackage{amstext}
\usepackage{upgreek}
\usepackage{multicol}
\usepackage{indentfirst}
\usepackage{graphicx}
\usepackage{paralist}
\usepackage{multirow}
\usepackage{tabularx}
\usepackage{cite}
\usepackage{citesort}
\usepackage{booktabs}
\usepackage{subfigure}
\usepackage{color}
\IEEEoverridecommandlockouts
\allowdisplaybreaks

\newtheorem{corollary}{\bf Corollary}
\newtheorem{proposition}{\bf Proposition}

\usepackage{accents}

\usepackage{flushend}
\allowdisplaybreaks
\begin{document}
\title{When Exploiting Individual User Preference Is Beneficial for Caching at Base Stations}
\author{
	\IEEEauthorblockN{{\large Dong Liu$^\star$, Chenyang Yang$^\star$, and Victor C.M. Leung$^\dagger$ }}\\
	\vspace{-3mm}
	\IEEEauthorblockA{$^\star$School of Electronic and Information Engineering, Beihang University, Beijing, China}
	\IEEEauthorblockA{$^\dagger$Department of Electrical and Computer Engineering, the University of British Columbia, Vancouver, BC, Canada \\
		Email: \{dliu, cyyang\}@buaa.edu.cn, vleung@ece.ubc.ca}
	\thanks{This work was supported in part by National Natural Science Foundation of China (NSFC) under Grant 61731002, 61671036, and 61429101.}
	\vspace{-5mm}
}
	\maketitle
\begin{abstract}
Most of prior works optimize caching policies based on the following assumptions: 1) every user initiates request according to \emph{content popularity}, 2) all users are with the same activity level, and 3) users are uniformly located in the considered region. In practice, these assumptions are often not true.
 In this paper, we explore the benefit of optimizing caching policies for base stations by exploiting \emph{user preference} considering the spatial locality and different activity level of users. We obtain optimal caching policies, respectively minimizing the download delay averaged over all file requests and user locations in the network (namely \emph{network average delay}), and minimizing the maximal weighted download delay averaged over the file requests and location of each user (namely \emph{maximal weighted user average delay}), as well as minimizing the weighted sum of both. The analysis and simulation results show that exploiting heterogeneous user preference and activity level can improve user fairness, and can also improve network performance when users are with spatial locality.
\end{abstract}

\section{Introduction}
By caching popular contents at base stations (BSs), user experience, network throughput, and energy efficiency can be improved remarkably \cite{bastug2014living,liu2016energy,liu2016caching}.

To achieve high performance with limited cache size at wireless edge, optimizing proactive caching is critical by harnessing the knowledge of which and where the contents will be requested. In an early work \cite{femtocachingTIT}, caching policy was optimized to minimize the average download delay assuming that the exact location where each user sends the file request is known \emph{a priori}. Considering the uncertainty in where the users will send requests, a probabilistic caching policy maximizing the cache-hit probability was proposed in \cite{Blaszczyszyn2015optimal}. In the literature of wireless caching, the knowledge of which contents will be demanded is commonly interpreted as content popularity. As a result,
most of prior works optimize caching policies based on \emph{content popularity} \cite{femtocachingTIT,Blaszczyszyn2015optimal,cuiying,chenzhen,cuiying,xiuhua,TMX,zhangrui}.

However, as a demand statistic of multiple users, content popularity cannot reflect the demand statistic of each individual user. In fact, global content popularity observed at a large aggregation point (say a content server) cannot reflect local content popularity observed in a small region (say a campus \cite{zink2009characteristics} or a cell \cite{ahlehagh2014video}), not to mention the preference of each user. These existing works implicitly assume that the preferences are identical among users in a region \cite{femtocachingTIT,Blaszczyszyn2015optimal,cuiying,chenzhen,cuiying,xiuhua,TMX} or in a social group \cite{zhangrui} and are equal to the content popularity. This inevitably  degrades the caching gain, since the assumption is not true in practice.

In real-world networks, user preferences are heterogeneous, which can be learned from collaborative filtering (CF) based on users' rating or request history \cite{ekstrand2011collaborative}. By assuming user preferences as Zipf distributions with different ranks, caching policy  was optimized to minimize the average download delay in \cite{liujuan}. Yet the user locations were assumed unchanged during the period of content placement and content delivery and all the users were assumed to have identical activity level. In practice, the location of mobile users is neither known in advance as assumed in \cite{bastug2014living,femtocachingTIT,chenzhen,liujuan}, nor completely unknown (hence randomly distributed throughout the network) as assumed in \cite{Blaszczyszyn2015optimal,cuiying,xiuhua,TMX}. The data measured from mobile connections in \cite{traffic,traffic2} showed that more than one third of the users visit only one cell and over 90\% of the users travel across less than 10 BSs  in one day, which indicates strong spatial locality of users. This suggests that the probability that a user is located in a cell when sending file request can be learned from the request history.  Moreover, the activity level of users is highly heterogeneous, e.g., about 80\% of the daily network traffic is generated by only 20\% of the users \cite{traffic2}.

In this paper, we analyze when optimizing caching policy with individual user preference is beneficial. Taking the spatial locality and different activity levels of users into account, we first derive the average delay for each user. We then minimize the network average delay, and show that exploiting user preference can improve network performance when users send requests with high probabilities in some cells. Noticing that user fairness issue appears when different users prefer the BSs to cache different files due to diverse user preference, we minimize the maximal weighted average delay among all the users, and show that caching policy can improve user fairness when user preferences are exploited.
The rest of the paper is organized as follows. In section II, we first introduce the system model, caching policy, and then connect content popularity with user preference. In section III, we optimize the caching policy with user preference, show when using user preference is beneficial, and use a toy example to help understand the impact of user preference heterogeneity. In section IV, simulation results are provided. Section V concludes the paper.

\section{System Model and User Demand Statistics}
\begin{figure}[!htb]
	\centering
	\includegraphics[width=0.3\textwidth]{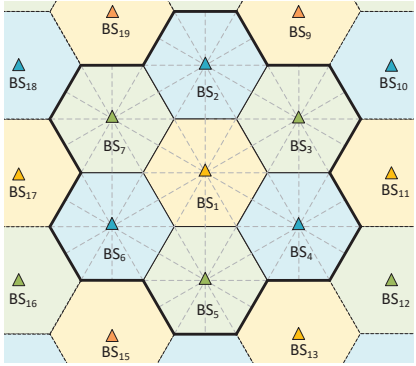}
	\caption{Layout of the cache-enabled network. The considered region are surrounded by solid line. In this example, $N_b = 7$.} \label{fig:layout}
	\vspace{-5mm}
\end{figure}

We consider a cache-enabled wireless network with cell radius $D$, where $N_b$ BSs serve $N_u$ users.  Each BS is equipped with $N_t$ antennas and a cache with size $N_c$, and is connected to the core network via backhaul. The content library consists of $N_f$ files each with size $F$ that all the users in the considered region may request. Each user is allowed to associate with one of the three nearest BSs (called \emph{neighboring BS set}) to download the requested file in order to increase the cache-hit probability. For example, when a user is located in the shaded area of Fig.~\ref{fig:layout}, it can associate with BS$_1$, BS$_2$ or BS$_3$, where BS$_1$ is called the local BS of the user.\footnote{The framework can be extended to neighboring BS sets with any number of BSs. We choose three only for illustration.} To avoid strong inter-cell interference inside the neighboring BS sets, the BSs within the neighboring BSs set use different frequency bands as shown in Fig. \ref{fig:layout}. To reflect the spatial locality of each user, we denote $\mathbf{A} = [a_{uj}]_{Nu\times N_b}$ as the location probability matrix, where $a_{uj}$ is the probability that the $u$th user is located in the $j$th cell when it sends file request. Since the exact location of users in a cell is hard to predict, we assume that the user is uniformly located within a cell when it is located in the cell.

\subsection{Caching Policy and Download Delay}
To achieve better performance, we employ coded caching strategy \cite{femtocachingTIT,TMX} where each file is encoded by rateless maximum distance separable coding so that a file can be retrieved by a user when $F$ bits of the requested file is received by the user. Denote $c_{bf}$ $(0\leq c_{bf} \leq 1)$ as the fraction of the $f$th file cached at the $b$th BS, and $b_u^l$ as the $l$th nearest BS of the $u$th user when the user receives file at the location of $\mathbf{x}_u = (x_{u1}, x_{u2})$.

When $\sum_{l=1}^{k - 1}c_{b_{u}^lf} < 1$ and $\sum_{l=1}^{k}c_{b_{u}^lf} \geq 1$, the $u$th user needs to receive the $f$th file from the $1$st, $\cdots$, $k$th nearest BSs\footnote{To unify the expression, we refer the backhaul to as the 4th nearest ``BS''.} successively to retrieve the complete file. Then, the delay of the $u$th user that locating at $\mathbf{x}_u$ and downloading the $f$th file averaged over small-scale fading can be expressed as a piecewise function
\begin{equation}
t^{f}_{u}(\mathbf{x}_u) =  \left\{\begin{array}{ll}
t^{f1}_{u}(\mathbf{x}_u),  & c_{b_{u}^1f} \geq 1\\
t^{f2}_{u}(\mathbf{x}_u),  & \sum_{l=1}^{1}c_{b_{u}^lf} < 1~\text{and}~ \sum_{l=1}^{2}c_{b_{u}^lf} \geq 1\\
t^{f3}_{u}(\mathbf{x}_u),  & \sum_{l=1}^{2}c_{b_{u}^lf} < 1~\text{and}~ \sum_{l=1}^{3}c_{b_{u}^lf} \geq 1\\

t^{f4}_{u}(\mathbf{x}_u), & \sum_{l=1}^{3}c_{b_{u}^lf} < 1
\end{array}
\right. \!\!\!\! \label{eqn:piecewise}
\end{equation}
where
\begin{equation}
t^{fk}_{u}(\mathbf{x}_u) = F \sum_{l=1}^{k - 1}c_{b_{u}^lf} \tau_{ub_{u}^l}(\mathbf{x}_u)  + F\Big(1-\sum_{l=1}^{k-1} c_{b_{u}^lf} \Big) \tau_{ub_{u}^k}(\mathbf{x}_u) \label{eqn:tufk}
\end{equation}
and $\tau_{ub_u^l}(\mathbf{x}_u)$ is the per-bit download delay of the $u$th user when downloading from its $l$th nearest BS averaged over small-scale fading.

We assume block Rayleigh fading channel, which is constant in each block and independently and identically distributed among blocks. Then, the per-bit download delay can be derived as \cite{liujuan}
\begin{equation}
\tau_{ub_u^l}(\mathbf{x}_u) = \frac{1}{\bar R_{ub_u^l}(\mathbf{x}_u)}  \label{eqn:tau}
\end{equation}
where $\bar R_{ub}(\mathbf{x}_u)$ is the achievable rate averaged over small-scale fading for the $u$th user that downloading from the $b$th BS. To unify the expression, we denote the download delay when the $u$th user downloading from the backhaul as $\tau_{ub_u^4}$. Since cache is intended for networks with stringent capacity backhaul \cite{bastug2014living}, we assume that the download delay is limited by the backhaul bandwidth when the user downloads file from the backhaul. Then, we have $\tau_{ub_u^4} = \frac{1}{C_{{\rm bh}, u}}$, where $C_{{\rm bh, u}}$ is the backhaul bandwidth for the $u$th user.

To emphasize how to optimize caching policy exploiting user preference, we assume that each BS serves $N_t$ users in the same time-frequency resource by zero-forcing beamforming with equal power allocation, then the average achievable rate can be expressed as
\begin{equation}
\bar R_{ub}(\mathbf{x}_u) \!= \mathbb{E}_{h}\! \left[W_u \log_2\left( \!1\! +\! \tfrac{\frac{P_t}{N_t} h_{ub}r_{ub}^{-\alpha}}{\sum_{b' \!\in \Phi(b), b' \neq b} P_t h_{ub'} r_{ub'}^{-\alpha} + \sigma^2 }\! \right) \!\right]\!\! \label{eqn:Rxy}
\end{equation}
where $W_u$ is the transmission bandwidth for the $u$th user, $P_t$ is the transmit power of each BS, $h_{ub}$ is the equivalent channel gain (including channel coefficient and beamforming) from the $b$th BS to the $u$th user, $r_{ub} = ||\mathbf{x}_u - \mathbf{x}_b||$ is the distance between the $u$th user and the $b$th BS, $\alpha$ is the pathloss exponent, $\Phi_b$ denotes the set of BSs that share the same frequency with the $b$th BS, $\sigma^2$ is the noise power, $\frac{P_t}{N_t} h_{ub}r_{ub}^{-\alpha}$ and  $\sum_{b' \!\in \Phi(b), b' \neq b} P_t h_{ub'} r_{ub'}^{-\alpha}$ are the signal power and interference power, respectively.

\subsection{Content Popularity and User Preference}
We denote $\mathbf{p} = [p_1, \cdots, p_{N_f}]$ as \emph{global content popularity}, where $p_f$ is the probability that the $f$th file is requested by all users in the considered region. We denote $p_{f|j}$ as the \emph{local content popularity} of the $f$th file in the $j$th cell, which is the probability that the $f$th file is requested by all users in the $j$th cell and reflects the user demands observed within a cell.

We denote $\mathbf{Q} = [\mathbf{q}_1^T, \cdots, \mathbf q_{N_u}^T]^T$ as \emph{user preference} matrix, where $\mathbf{q}_u = [q_{1|u}, \cdots, q_{N_f|u}]$ is the preference of the $u$th user and $q_{f|u} \in [0,1]$ is the conditional probability that the $u$th user requests the $f$th file given that it requests a file. User preference reflects the demands of each individual user.

Based on the law of total probability, the global content popularity can be connected with user preference as
\begin{equation}
p_{f}  = \sum_{u=1}^{N_u} s_uq_{f|u} \triangleq \sum_{u=1}^{N_u} q_{uf} \label{eqn:relation}
\end{equation}
where $s_u$ is the probability that the request is sent from the $u$th user, which reflects the \emph{activity level} of the user, and $q_{uf}$ is the joint probability that the requested file is the $f$th file and the request is sent from the $u$th user. We denote $\mathbf{s} = [s_1, \cdots, s_{N_u}]$ as the user activity level vector.

Further considering the user location probability $\mathbf{A}$, the local content popularity of the $f$th file in the $j$th cell can be connected with user preference as
\begin{equation}
p_{f|j} = \frac{\sum_{u = 1}^{N_u}a_{uj} s_u q_{f|u}}{\sum_{f = 1}^{N_f}\sum_{u = 1}^{N_u}a_{uj} s_u q_{f|u}} = \frac{\sum_{u = 1}^{N_u}a_{uj} q_{uf}}{\sum_{u = 1}^{N_u}a_{uj}s_u} \label{eqn:local}
\end{equation}

%

Both $\mathbf{Q}$ and $\mathbf{s}$ can be learned by CF at a service gateway \cite{CBQ,bigdata}, which are assumed perfect in the following analysis.

\section{Caching Policy Optimization With User Preference}
In practice, the exact location where each user sends the file request is unknown in advance when optimizing the caching policy. Therefore, we first derive the delay of each user averaged over its possible locations and file requests.

To derive the user average delay, we divide each cell into 12 sectors as shown in Fig.~\ref{fig:layout}. In this way, the $l$th nearest BS of the $u$th user, i.e., $b_u^l$, does not depend on $\mathbf{x}_u$ any more given that the user is located in the $i$th sector of the $j$th cell. Then, based on the law of total expectation, the average delay of the $u$th user can be obtained by the following proposition

\begin{proposition}
	The download delay of the $u$th user averaged over all its possible requests and locations is
\begin{equation}
\bar t_u \!= \!\sum_{j=1}^{N_b}\sum_{i=1}^{12} \sum_{f=1}^{N_f}\frac{a_{uj}q_{f|u}}{12}\!\!\!\max_{k=1,\cdots,4}\!\Big\{\! F \bar\tau_{uk} - F\!\sum_{l=1}^{k - 1}c_{b_{ij}^lf}\! \left(\bar\tau_{uk} - \bar\tau_{ul}\right)\!\!\Big\} \label{eqn:bartu}
\end{equation}
where $\bar \tau_{ul} \approx \frac{2\sqrt{3}}{W_uD^2} \int_{0}^{D} \int_{0}^{\frac{x_{2}}{\sqrt{3}}}  \frac{1}{\log_2 \frac{k_x}{k_y} +\frac{1}{\ln 2}(\psi(\theta_x)- \psi(\theta_y))}  {\rm d}x_{1} {\rm d}{x_{2}}$ is the  per-bit download delay of the $u$th user when downloading from the $l$th nearest BS averaged over the $u$th user's location, $k_x, k_y, \theta_x$ and $\theta_y$ are given in the Appendix.
\end{proposition}

\begin{IEEEproof}
	See Appendix.
\end{IEEEproof}

\subsection{Caching Policy Optimization}
Network average delay is the delay averaged over the requests of all the users in the considered region. This is a performance metric from the network perspective and is widely used in literature~\cite{femtocachingTIT,zhangrui,liujuan}, which can be expressed as $ T = \sum_{u=1}^{N_u} s_u \bar t_u$.

To capture user fairness, we consider the weighted user average delay $\max\limits_{u=1,\cdots, N_u}\{w_u\bar t_u\}$. Considering that the users with more file requests will suffer more if they have longer delay, we can set $w_u$ as an increasing function of the user activity level $s_u$. As an illustration, we set $w_u = N_us_u$ in the sequel. Then, the weighted user average delay can be expressed as $w_u\bar t_u=N_u s_u \bar t_u$.

To improve both network performance and user fairness, we formulate the following general optimization framework minimizing the weighted sum of these two metrics as
\begin{subequations}
	\begin{align}
	\min_{c_{bf}} ~~& (1-\eta) T + \eta\max_{u=1,\cdots,N_u}\left\{ N_u s_u \bar t_u \right\} \\
	s.t. ~~&\sum_{f=1}^{N_f} c_{bf} \leq N_c,~ \forall b \label{eqn:con1}\\
	& 0\leq c_{bf} \leq 1,~ \forall f, b \label{eqn:con2}
	\end{align} \label{eqn:P0}
\end{subequations}
By changing the value of $\eta$ from $0$ to $1$, we can obtain the caching policy from minimizing the network average delay (refer to as \emph{Problem 1}) to minimizing the maximal weighted user average delay (refer to as \emph{Problem 2}). By introducing auxiliary variables $\mu_{ij}^{uf}$ and $\nu$, which are upper bounds of $\{ \bar\tau_{ul} - \sum_{l=1}^{k - 1}c_{b_{ij}^lf} (\bar\tau_{uk} - \bar\tau_{ul}) \}_{k = 1, \cdots, 4}$ and $\{N_u s_u \bar t_u\}_{u = 1, \cdots, N_u}$, respectively, we can convert the problem equivalently into
\begin{subequations}
	\begin{align}
	\!\!\!\!\!\!\min_{c_{bf}, \mu_{ij}^{uf}, \nu} & (1-\eta) \frac{F}{12}\sum_{j=1}^{N_b}\sum_{i=1}^{12} \sum_{f=1}^{N_f}\left(\sum_{u=1}^{N_u}a_{uj}q_{uf} \right)\mu_{ij}^{uf} + \eta\nu \label{eqn:obj}\\
	s.t. ~~&\bar\tau_{uk} - \sum_{l=1}^{k - 1}c_{b_{ij}^lf} (\bar\tau_{uk}- \bar\tau_{ul})  \leq \mu_{ij}^{uf}, \forall~ i,j,u,f,k \\
	& N_u\sum_{j=1}^{N_b}\sum_{i=1}^{12} \sum_{f=1}^{N_f}a_{uj}q_{uf} \mu_{ij}^{uf} \leq \nu ,~ \forall u \\
&\sum_{f=1}^{N_f} c_{bf} \leq N_c,~ \forall b\\
& 0\leq c_{bf} \leq 1,~ \forall f, b
	\end{align}
\end{subequations}
which is a linear programming problem and can be solved by interior point method \cite{boyd2004convex}. We refer the optimal caching policies for Problem 1 and Problem 2 to as \emph{Policy} 1 and \emph{Policy} 2, respectively.

\subsection{Analysis for Special Cases}
Since transmission and caching resource allocation operated in very different time-scales, to focus on the difference brought by exploiting user preference, we consider the special cases where transmission resources are identical for each user  (i.e., $\bar \tau_{1l} = \cdots = \bar \tau_{N_ul} \triangleq \bar \tau_l$) in the following. Depending on whether the coverage areas of BSs are overlapped, we analyze Policy 1 and Policy 2 in two scenarios.

No matter the coverage of adjacent BSs overlap or not, we can obtain Corollaries 1 and 2 in the following.
\begin{corollary}
	When each user sends request in uniform-distributed locations throughout the network, exploiting user preference cannot improve network average delay.
\end{corollary}
\begin{IEEEproof}
	In this case, we have $a_{u1} = \cdots = a_{uN_b} = \frac{1}{N_b}$, $\bar \tau_{1l} = \cdots = \bar \tau_{N_ul}$, and $\mu_{ij}^{1f} = \cdots = \mu_{ij}^{N_uf} \triangleq \mu_{ij}^{f}$. Then, the first term in \eqref{eqn:obj} can be rewritten as
	\begin{equation}
	\frac{F}{12 N_b}\sum_{j=1}^{N_b}\sum_{i=1}^{12} \sum_{f=1}^{N_f}\Big(\!\sum_{u=1}^{N_u}q_{uf} \!\Big)\mu_{ij}^{f}  =  \frac{F}{12 N_b}\sum_{j=1}^{N_b}\sum_{i=1}^{12} \sum_{f=1}^{N_f}p_f\mu_{ij}^{f}  \nonumber
	\end{equation}
	where we use the relation in \eqref{eqn:relation}. We can see that the network average delay only depends on global content popularity $p_f$.
\end{IEEEproof}

\begin{corollary}
	When the location probabilities and preferences are identical for all users, Policies 1 and 2 are identical.
\end{corollary}
\begin{IEEEproof}
	In this case, since $\bar \tau_{1l} = \cdots = \bar \tau_{N_ul}$, $a_{1j} = \cdots = a_{N_uj}$ for all $j$, and $q_{f|1} = \cdots = q_{f|N_u}$ for all $f$, we can see from \eqref{eqn:bartu} that the average delay of each user is identical, i.e. , $\bar t_1 = \cdots = \bar t_{N_u} \triangleq \bar t$. Then,  both Problem 1 and Problem 2 are equivalent to minimizing $\bar t$.
\end{IEEEproof}

From the corollaries  we can conclude that if the transmission resources are identical for all users,
the gain of exploiting user preference in terms of network average delay will vanish without user spatial locality.
If location distributions and preferences are further identical for all users, the maximal weighted user average delay can be minimized by simply minimizing the average network delay.

In sparse networks where the coverage of adjacent BSs do not overlap, the average delay of the $u$th user degenerates into
\begin{equation}
\bar t_u = F \sum_{j=1}^{N_b}  \sum_{f = 1}^{N_f} a_{uj} q_{f|u}(c_{jf} \bar \tau_{1} +  (1-c_{jf}) \bar \tau_{4}) \label{eqn:degen}
\end{equation}

\begin{corollary}
	When the coverage areas of BSs are non-overlapped, Policy 1 is to let each BS cache the most popular files according to local content popularity.
\end{corollary}
\begin{IEEEproof}
In this case, considering \eqref{eqn:local} and \eqref{eqn:degen}, we can obtain $T = F\sum_{j=1}^{N_b}(\sum_{u = 1}^{N_u}a_{uj}s_u)( \sum_{f=1}^{N_f} p_{f|j} (c_{jf}\bar \tau_1 + (1 - c_{jf})\bar \tau_4)  )$. Then, minimizing $T$ is equivalent to minimizing $\sum_{f=1}^{N_f} p_{f|j} (c_{jf}\bar \tau_1 + (1 - c_{jf})\bar\tau_4)$ for each cell, $j=1,\cdots, N_b$, which can be rewritten as $\sum_{f=1}^{N_f} p_{f|j}\bar\tau_4  - \sum_{f=1}^{N_f} p_{f|j} c_{jf} (\bar  \tau_4 - \bar \tau_1 )$. Since $\bar \tau_4 > \bar \tau_1$, it is easy to see that the optimal caching policy is to let each cell cache the $N_c$ complete files with the highest values of $p_{f|j}$.
\end{IEEEproof}

\begin{corollary}
	When the coverage areas of BSs are non-overlapped and user preference is identical, Policy 1 and Policy 2 are the same.
\end{corollary}
\begin{IEEEproof}
In this case, $\bar t_u = a_{uj} F \sum_{j=1}^{N_b}  \sum_{f = 1}^{N_f} p_f (c_{jf} \bar \tau_{1} +  (1-c_{jf}) \bar \tau_{4})$. Then, both minimizing $\max_{u=1,\cdots,N_u}\{N_u s_u \bar t_u\} $ and minimizing $T = \sum_{u=1}^{N_u} s_u \bar t_u$ are equivalent to minimizing $\sum_{j=1}^{N_b}  \sum_{f = 1}^{N_f} p_f (c_{jf} \bar \tau_{1} +  (1-c_{jf}) \bar \tau_{4})$. Therefore, Policy 1 and Policy 2 are the same.
\end{IEEEproof}

From Corollary 3 and Corollary 4, we can conclude that if the transmission resources are identical for users and the cells are not overlapped, using local content popularity will be enough to obtain the minimal network average delay as used in \cite{ahlehagh2014video,bigdata}. Otherwise, user preference should be exploited to minimize the network average delay.
If user preference is further identical, the maximal weighted user average delay can be minimized by simply minimizing the average network delay.
Otherwise, caching policies should be designed more sophisticatedly to address user fairness issue.

\subsection{Numerical Examples}
\begin{figure}[!htb]
	\vspace{-2mm}
	\centering
		\label{fig:example2} 
		\includegraphics[height=0.12\textwidth]{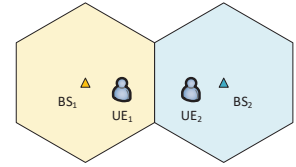}
	\caption{A toy example, $N_b = N_u = 2$. The total number of files is $N_f = 3$ with size $F = 1$ and each BS can cache $N_c = 1$ file. Global content popularity is $\mathbf p = [0.46, 0.30, 0.24]$ and activity level is $\mathbf s = [0.6, 0.4]$.}
	\label{fig:example} 
\end{figure}

To understand the behavior of Policy 1 and Policy 2, and analyze the impact of heterogeneous user preference, we present a simple numerical example as shown in Fig. \ref{fig:example}.

Suppose each user can either associate with BS$_1$ or BS$_2$ to download files, and the average per-bit download delay when downloading from the nearest BS, second nearest BS and the backhaul is $[\bar \tau_{u1}, \bar \tau_{u2}, \bar \tau_{u4}] = [1, 2, 3]$ for both user equipments (UEs).  We compare two cases with homogeneous and heterogeneous user preference, respectively, where both $\mathbf Q^{\rm hom}$  and $\mathbf Q^{\rm het}$ satisfy \eqref{eqn:relation} with given $\mathbf p$ and $\mathbf s$.
\begin{equation}
\mathbf{Q}^{\rm hom} =\begin{bmatrix}
0.46 & 0.30 & 0.24 \\
0.46 & 0.30 & 0.24
\end{bmatrix}
\!, ~
\mathbf{Q}^{\rm het} = \begin{bmatrix}
0.75 & 0.25 & 0 \\
0.02 & 0.38 & 0.60
\end{bmatrix} \nonumber
\end{equation}

For homogeneous user preference $\mathbf{Q}^{\rm hom}$, we can obtain the results of Policy 1 $\mathbf{C}_1$ and the minimized network average delay $T^*$, the results of Policy 2 $\mathbf{C}_2$ and the minimized maximal weighted user average delay $\max\{N_u s_1\bar t_1^{\dagger}, N_u s_2\bar t_2^{\dagger}\}$ as
\begin{align}
& \mathbf{C}_1 =\begin{bmatrix}
1 & 0 & 0 \\
0 & 1 & 0
\end{bmatrix},   ~T^* = s_1 \bar t_1^* + s_2 \bar t_2^*  =1.07 + 0.77 = 1.84 \nonumber \\
& \mathbf{C}_2 =\begin{bmatrix}
1 & 0 & 0 \\
0 & 1 & 0
\end{bmatrix}, ~\max \{N_u s_1\bar t_1^{\dagger}, N_u s_2\bar t_2^{\dagger}\} = 2.13 \nonumber
\end{align}
In this case, the \emph{cache interests} of both users are exactly the opposite, i.e., UE$_1$ prefers its local BS (i.e., BS$_1$) to cache its most preferable file (i.e., file $1$) and its neighboring BS (i.e., BS$_2$) to cache its second preferable file (i.e., file $2$), while UE$_2$ prefers BS$_2$ to cache file $1$ and BS$_1$ to cache file $2$ according to its own preference. Since UE$_1$ has higher activity level, both Policies 1 and 2 let BSs cache file according to UE$_1$'s cache interest and $\mathbf{C}_1 = \mathbf{C}_2$, which agrees with Corollary 2.

For heterogeneous user preference $\mathbf{Q}^{\rm het}$, we can obtain
\begin{align}
& \mathbf{C}_1 =\begin{bmatrix}
1 & 0 & 0 \\
0 & 0 & 1
\end{bmatrix},   ~T^* = s_1 \bar t_1^* + s_2 \bar t_2^*  =0.90 +0.71 = 1.61 \nonumber \\
& \mathbf{C}_2 =\begin{bmatrix}
1 & 0 & 0 \\
0 & 0.57 & 0.43
\end{bmatrix},  ~\max \{N_u s_1\bar t_1^{\dagger}, N_u s_2\bar t_2^{\dagger}\} = 1.63 \nonumber
\end{align}
In this case, UE$_1$ prefers BS$_1$ to cache file $1$ and BS$_2$ to cache file $2$, while UE$_2$ prefers BS$_2$ to cache file $3$ and BS$_1$ to cache file $2$. As a result, Policy 1 lets each BS cache the most preferable file of its local user, i.e. BS$_1$ caches file $1$ and BS$_2$ caches file $3$. As the user with higher activity level, UE$_1$ has the maximal weighted average delay (i.e., $0.90 > 0.71$). Hence, Policy 2 is more prone to let BSs cache the files preferred by UE$_1$, i.e., let BS$_2$ cache $0.57$ part of file $2$ and $0.43$ part of file $3$. We can see that both the average network delay and the maximal weighted average delay decrease compared to the case with homogeneous user preference, which can be explained from the following different perspective.

When user preference become heterogeneous, the most preferable files of users located in different cells differ. File diversity (i.e., caching different files at different BSs) can be naturally achieved by letting each BS cache the most preferable file of its local user, which increases the cache-hit probability. On the contrary, when user preference is identical, there will be no file diversity if each BS caches the most preferable file of its local user, and to achieve file diversity, the cache interest of UE$_2$ has to be sacrificed.

With given content popularity, the skewness of both users' preferences increase (i.e., the shape of probability distribution $[0.75, 0.25, 0]$ and $[0.02, 0.38, 0.60]$ are more ``\emph{peaky}" than $[0.46, 0.30, 0.24]$) when user preferences are less similar, which means that the file requests of users become less uncertain. Analogously to the widely recognized result that the performance of content popularity based caching policies improves with the skewness of popularity, the performance of user preference based caching policies improves with the skewness of user preference.
\section{Simulation Results}
In this section, we compare the performance of the proposed caching policies with prior works that are based on content popularity, and analyze the impact of various factors by simulation.

We consider $N_b = 7$ cells each with radius $D = 250$ m as shown in Fig. \ref{fig:layout}, and $N_u = 100$ users. Each BS is with four antennas and with transmit power $46$ dBm. The pathloss is modeled as $35.5+37.6\log_{10}(r_{ub})$. The backhaul bandwidth and the downlink transmission bandwidth for each user are set as $C_{{\rm bh},u} = 2$ Mbps and $W_u = 5$ MHz, respectively. The probability distribution for the users located in different cells when sending requests is modeled as Zipf distribution with skewness parameter $\delta_a = 1$ based on the measured data in \cite{traffic}. To analyze the impacts of user preference and activity level and fairly compare with prior works, we generate user preferences satisfying the relation in \eqref{eqn:relation} with different level of cosine similarity as defined in \cite{CBQ}
. To reduce simulation time, we consider $N_f = 100$ files in total each with size of $F = 30$ MB. We assume that each BS can cache 10\% of the total files, i.e., $N_c = 10$. The global content popularity and the activity level are modeled as Zipf distribution with the skewness parameter $\delta_p = 0.6$ and $\delta_s = 0.4$, respectively.

The following baselines are compared with Policy 1 and Policy 2, where the activity levels and user preferences are implicitly assumed identical when designing caching policies for baselines 1) and 3):
\begin{enumerate}
	\item {\em``Global Pop"}: Each BS caches the $N_c$ most popular files according to the global content popularity $p_f$.
	\item {\em``Local Pop"}: Each BS caches the $N_c$ most popular files according to the local content popularity within its cell $p_{f|j}$ given by \eqref{eqn:local}. This is the method used in \cite{ahlehagh2014video,bigdata}.
	\item {\em ``Femtocaching (Pop)"}: This is the caching policy proposed in \cite{femtocachingTIT} minimizing the network average delay, which is based on global content popularity assuming that user location is fixed.
	\item {\em``Femtocaching (Pref)"}: We modify the caching policy in \cite{femtocachingTIT} to exploit user preference by simply replacing the global content popularity $p_f$ by user preference $q_{f|u}$.
\end{enumerate}

\begin{figure}[!htb]
	\vspace{-4mm}
	\centering
	\subfigure[Network performance]{
		\label{fig:T_vs_cos} 
		\includegraphics[width=0.23\textwidth]{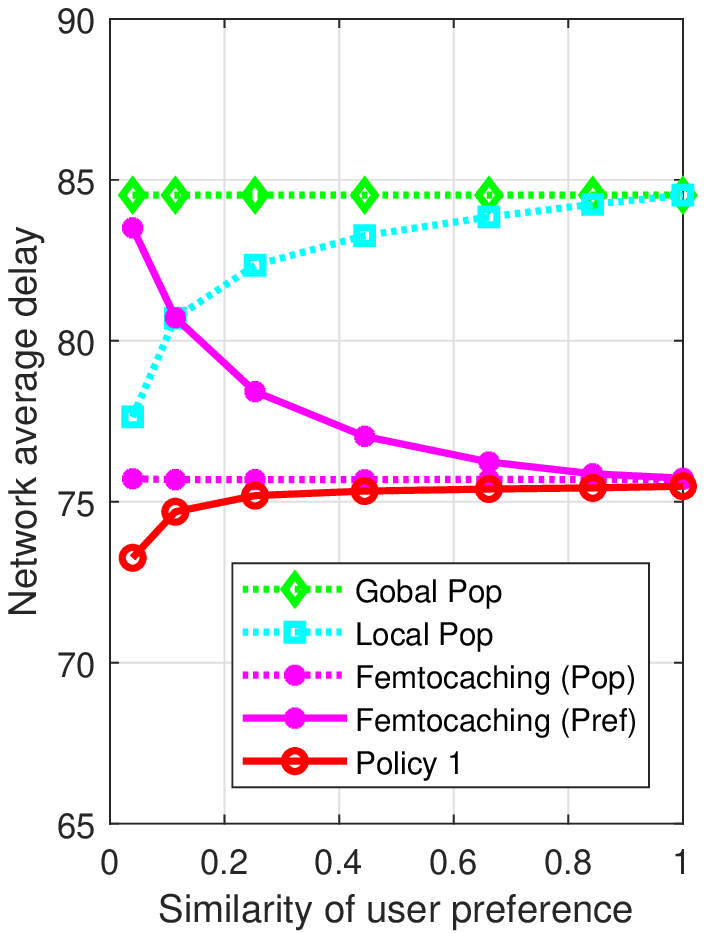}}
	\subfigure[User fairness]{
		\label{fig:max_vs_cos} 
		\includegraphics[width=0.23\textwidth]{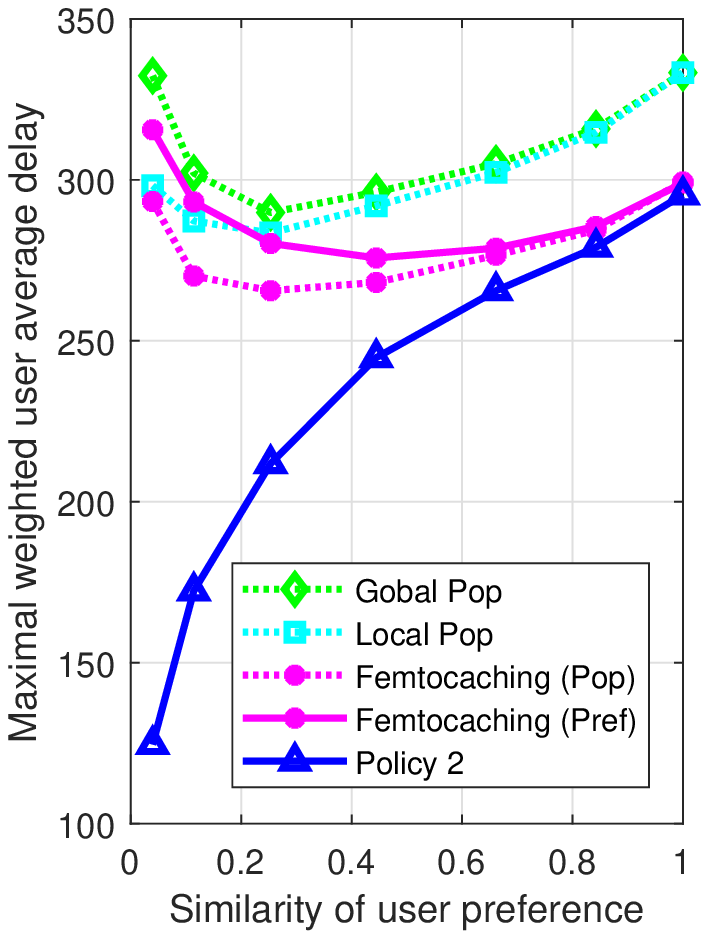}}
		\vspace{-1mm}
	\caption{Impact of user preference similarity. }
	\label{fig:cos} 
	\vspace{-1mm}
\end{figure}

In Fig.~\ref{fig:T_vs_cos}, we show the impact of user preference
similarity on the network average delay (in seconds). It is shown that ``Local Pop" can reduce  network average delay compared with ``Global Pop" when user preferences are heterogeneous.  The network average delay of ``Femtocaching (Pref)" is even higher than that of ``Femtocaching (Pop)" when user preference is less similar. This is because ``femtocaching" method does not consider the uncertainty of user location, which has large impact when user preference is less similar. The network average delay of Policy 1 is the lowest as expected, which increases with the preference similarity. This coincides with the results of numerical example in Section III-B.

In Fig. \ref{fig:max_vs_cos}, we show the impact of user preference similarity on the maximal weighted user average delay (in seconds). We can see that Policy 2 can reduce $60\%$  of the maximal weighted user average delay compared with ``Global Pop". Similar to Fig.~\ref{fig:T_vs_cos}, the maximal weighed download delay of ``Femtocaching (Pref)" is higher than ``Femtocaching (Pop)". The maximal weighted user average delay of Policy 2 is  the lowest, which increases with the preference similarity. The explanations are similar to those for numerical results in Section III-B.

\begin{figure}[!htb]
	\vspace{-3mm}
	\centering
	\subfigure[Network performance]{
		\label{fig:T_vs_a} 
		\includegraphics[width=0.23\textwidth]{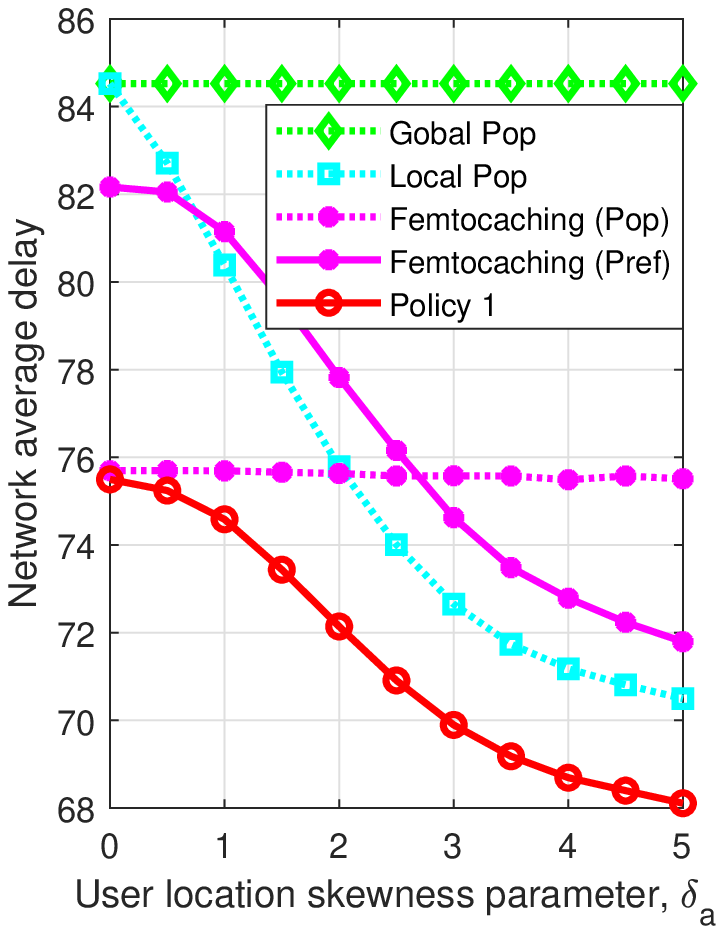}}
	\subfigure[User fairness]{
		\label{fig:max_vs_a} 
		\includegraphics[width=0.23\textwidth]{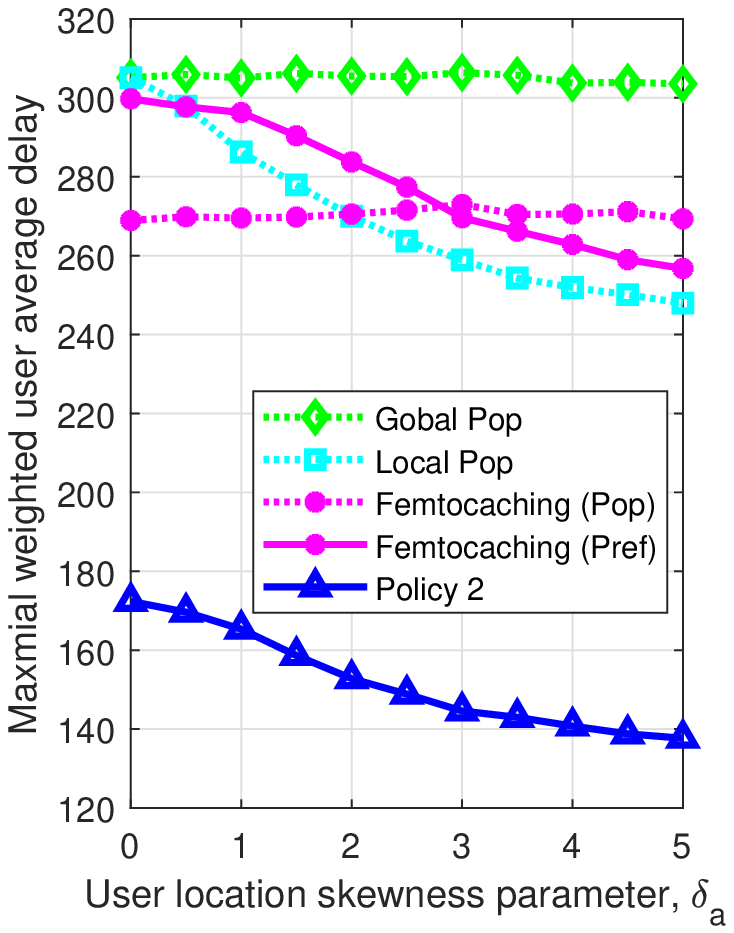}}
	\vspace{-2mm}
	\caption{Impact of  spatial locality of users. The similarity of user preference is 0.1.}
	\label{fig:a} 
	\vspace{-5mm}
\end{figure}

In Fig.~\ref{fig:a}, we show the impact of spatial locality on the two performance metrics. We can see that the benefit of exploiting user preference increases with spatial locality of users. Without spatial locality (i.e., $\delta_a = 0$), the network performance does not benefit from exploiting user preference while user fairness can still be improved when comparing Policies 1 and 2 with ``Femtocaching (Pop)", respectively.

\begin{figure}[!htb]
	\vspace{-4mm}
		\centering
	\includegraphics[height=0.28\textwidth]{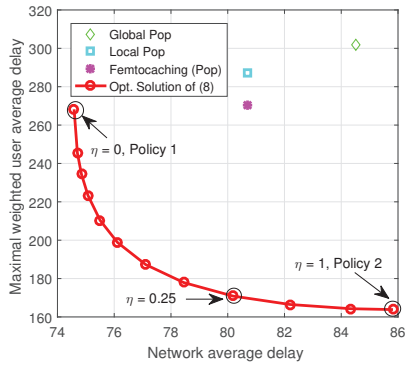}
	\vspace{-1mm}
	\caption{Tradeoff between performance and user fairness. The similarity of user preference is 0.1. }
		\label{fig:tradeoff} 
\end{figure}
In Fig. \ref{fig:tradeoff}, we show the tradeoff between network average download and maximal weighted average download delay by solving problem \eqref{eqn:P0} with different values of  $\eta$. It is shown that
when $\eta$ is set between 0 and 0.25, the optimal solution achieves lower network average delay and better user fairness than the baseline policies at the same time.

\section{Conclusion and Discussion}
In this paper, we strived to investigate when and how optimizing caching policy with user preference is beneficial. We showed that the network average delay can be reduced when users are with spatial locality, and user fairness can be improved when user preferences are heterogeneous. Simulation results showed that network performance and user fairness can even be improved at the same time compared with prior works by exploiting heterogeneous user preference and activity level with spatial locality. With given content popularity, the performance gain comes from the facts that cache-hit probability can be improved with less sacrifice of users' cache interests and user demands become less uncertain with more heterogeneous user preference.

It is worthy to mention that learning individual preference of a large number of users can be more computational complex than learning content popularity, and informing the predicted user preference to BSs may incur overhead. In practice, user preference can be learned  not very frequently (say each day) at a service gateway or even at a content server that has abundant computing resource. Nevertheless, to harness the benefit of user preference based caching policy, it is worthwhile to investigate how to reduce the complexity and overhead.
\appendix
Based on the law of total expectation, the average delay of the $u$th user can be expressed as
\begin{equation}
\bar t_u = \mathbb{E}_{f, \mathbf{x}_u} \!\!\left[t_u^f(\mathbf{x}_u)\right] = \sum_{f=1}^{N_f} \sum_{j=1}^{N_b}\sum_{i=1}^{12} \frac{a_{uj}}{12} \mathbb{E}_{\mathbf{x}_u}\!\!\left[t_u^f(\mathbf{x}_u)|ij\right] \label{eqn:Etuf}
\end{equation}
where $\mathbb{E}_{\mathbf{x}_u}\left[t_u^f(\mathbf{x}_u)| ij\right]$ is the average delay of the user conditioned on that it is located at the $i$th sector of the $j$th cell and requesting the $f$th file, and $\frac{a_{uj}}{12}$ is the probability that the $u$th user is located at the $i$th sector of the $j$th cell. Further considering \eqref{eqn:piecewise}, we can obtain
\begin{align}
\mathbb{E}_{\mathbf{x}_u}\left[t_u^f(\mathbf{x}_u) | ij\right] = \mathbb{E}_{\mathbf{x}_u}\left[t^{fk}_{u}(\mathbf{x}_u) |ij\right]
 \label{eqn:Epiecewise}
\end{align}
when $\sum_{l=1}^{k-1}c_{b_{ij}^lf} < 1$ and $\sum_{l=1}^{k}c_{b_{ij}^lf} \geq 1$, where $b_{ij}^l$ is the $l$th nearest BS when the user is located in the $i$th sector of the $j$th cell. From \eqref{eqn:tufk}, we can obtain
\begin{equation}
\mathbb{E}_{\mathbf{x}_u} \left[t^{fk}_{u}(\mathbf{x}_u) | ij\right]  = F \bar\tau_{uk} - F\sum_{l=1}^{k - 1}c_{b_{ij}^lf} (\bar\tau_{uk} - \bar\tau_{ul})
 \label{eqn:Etufk}
\end{equation}
where $\bar \tau_{ul} \triangleq \mathbb{E}_{\mathbf{x}_u} [\tau_{ub_{ij}^l}(\mathbf{x}_u) | ij]$ is the per-bit delay from the $l$th nearest BS averaged over user location given that the user is located in the $i$th sector of the $j$th cell.

Since the average delay increases with the distance between user and BS, we have $\bar \tau_{uk} > \bar \tau_{u(k-1)}$. Further considering the expressions of \eqref{eqn:Epiecewise} and \eqref{eqn:Etufk}, similar to the proof of \cite[Lemma 6]{femtocachingTIT}, we can rewrite \eqref{eqn:Epiecewise} as
\begin{equation}
\mathbb{E}_{\mathbf{x}_u}\left[t_u^f(\mathbf{x}_u) | ij\right] = \max_{k = 1,\cdots, 4}\{\mathbb{E}_{\mathbf{x}_u}\left[ t^{fk}_{u}(\mathbf{x}_u) | ij\right]\} \label{eqn:Etufij}
\end{equation}

Due to the symmetry of the network topology, $\bar \tau_{ul}$ does not depend on $i$ and $j$ but only depend on $l$ and $u$. Without loss of generality, we derive the average download delay from the three nearest BSs when the $u$th user located in the shadow area in Fig. \ref{fig:layout}. From \eqref{eqn:tau}, by taking the expectation over $\mathbf{x}_u$ within the shadow area, we have
\begin{equation}
 \bar \tau_{ul} = 2\sqrt{3} W_u^{-1}D^{-2}\int_{0}^{D}\int_{0}^{\frac{x_{u2}}{\sqrt{3}}}\bar R_{ul}(\mathbf{x}_u)^{-1} {\rm d}x_{u1} {\rm d}x_{u2} \label{eqn:int}
\end{equation}

Since the interference term in \eqref{eqn:Rxy} is a weighted sum of Gamma distributed random variables $h_{ub'} \sim \mathbb{G}(N_t, 1/N_t)$, with different values of weight $r_{ub'}^{-\alpha}$, $\bar R_{ul}(\mathbf{x}_u)$ has no closed-form expression and the computation requires a $|\Phi_l|$-fold numerical integration that is of high complexity. To reduce computational complexity, we obtain an approximate $\bar \tau_{ul}$ for high signal-to-noise ratio (SNR) region.

When $\frac{P_t}{\sigma^2} \to \infty$, we can neglect the impact of $\sigma^2$ and \eqref{eqn:Rxy} can be derived as
\begin{align}
\bar R_{ub} (\mathbf{x}_u)  & = W_u\mathbb{E}_{h}[\log_2  X  ]  - W_u \mathbb{E}_{h}[\log_2  Y] \nonumber \\
&\approx W_u\mathbb{E}_{h}[\log_2 \hat X  ]  - W_u \mathbb{E}_{h}[\log_2 \hat Y] \nonumber \\
& =  W_u \left(\log_2 \tfrac{k_x}{k_y}  + \tfrac{1}{\ln 2}(\psi(\theta_x) - \psi(\theta_x))\right) \label{eqn:approx}
\end{align}
where $X = h_{ub}r_{ub}^{-\alpha} + N_t\sum\nolimits_{b'\in \Phi(b), b'\neq b} h_{ub'} r_{ub'}^{-\alpha}$, $Y = N_t\sum\nolimits_{b'\in \Phi(b), b'\neq b} h_{ub'} r_{ub'}^{-\alpha}$, and we approximate $X$ and $Y$ as Gamma distributed random variables $\hat X \sim \mathbb{G}(k_x, \theta_x)$ and $\hat Y \sim \mathbb{G}(k_y, \theta_y)$, respectively, which is accurate as shown in \cite{gamma}. The last equation is from $\mathbb{E}[\ln \hat X] = \psi(k_x) + \ln(\theta_x)$. By matching the first two moments of $X$ and $\hat X$, we can obtain $k_x = \frac{\left(r_{ub}^{-\alpha} + N_t\sum_{b'\in \Phi(b), b'\neq b} r_{ub'}^{-\alpha}\right)^2}{r_{ub}^{-2\alpha} + N_t\sum_{b'\in \Phi(b), b'\neq b} r_{ub'}^{-2\alpha}} $, $\theta_x = \frac{r_{ub}^{-2\alpha} + N_t\sum_{b'\in \Phi(b), b'\neq b} r_{ub'}^{-2\alpha}}{r_{ub}^{-\alpha} + N_t\sum_{b'\in \Phi(b), b'\neq b} r_{ub'}^{-\alpha}}$, $k_y = \frac{N_t\left( \sum_{b'\in \Phi(b), b'\neq b} r_{ub'}^{-\alpha}\right)^2}{\sum_{b'\in \Phi(b), b'\neq b} r_{ub'}^{-2\alpha}} $, and $\theta_y =  \frac{\sum_{b'\in \Phi(b), b'\neq b} r_{ub'}^{-2\alpha}}{\sum_{b'\in \Phi(b), b'\neq b} r_{ub'}^{-\alpha}}$.

Then, by substituting \eqref{eqn:approx} into \eqref{eqn:int} and then into \eqref{eqn:Etufk} and further considering \eqref{eqn:Etufij} and \eqref{eqn:Etuf}, Proposition 1 can be proved.
\bibliographystyle{IEEEtran}
\bibliography{dongbib}

\begin{thebibliography}{10}
\providecommand{\url}[1]{#1}
\csname url@samestyle\endcsname
\providecommand{\newblock}{\relax}
\providecommand{\bibinfo}[2]{#2}
\providecommand{\BIBentrySTDinterwordspacing}{\spaceskip=0pt\relax}
\providecommand{\BIBentryALTinterwordstretchfactor}{4}
\providecommand{\BIBentryALTinterwordspacing}{\spaceskip=\fontdimen2\font plus
\BIBentryALTinterwordstretchfactor\fontdimen3\font minus
  \fontdimen4\font\relax}
\providecommand{\BIBforeignlanguage}[2]{{%
\expandafter\ifx\csname l@#1\endcsname\relax
\typeout{** WARNING: IEEEtran.bst: No hyphenation pattern has been}%
\typeout{** loaded for the language `#1'. Using the pattern for}%
\typeout{** the default language instead.}%
\else
\language=\csname l@#1\endcsname
\fi
#2}}
\providecommand{\BIBdecl}{\relax}
\BIBdecl

\bibitem{bastug2014living}
E.~Bastug, M.~Bennis, and M.~Debbah, ``Living on the edge: The role of
  proactive caching in {5G} wireless networks,'' \emph{IEEE Commun. Mag.},
  vol.~52, no.~8, pp. 82--89, Aug. 2014.

\bibitem{liu2016energy}
D.~Liu and C.~Yang, ``Energy efficiency of downlink networks with caching at
  base stations,'' \emph{IEEE J. Sel. Areas Commun.}, vol.~34, no.~4, pp.
  907--922, Apr. 2016.

\bibitem{liu2016caching}
D.~Liu, B.~Chen, C.~Yang, and A.~F. Molisch, ``Caching at the wireless edge:
  design aspects, challenges, and future directions,'' \emph{IEEE Commun.
  Mag.}, vol.~54, no.~9, pp. 22--28, Sept. 2016.

\bibitem{femtocachingTIT}
K.~Shanmugam, N.~Golrezaei, A.~G. Dimakis, A.~F. Molisch, and G.~Caire,
  ``Femtocaching: Wireless content delivery through distributed caching
  helpers,'' \emph{IEEE Trans. Inf. Theory}, vol.~59, no.~12, pp. 8402--8413,
  Dec 2013.

\bibitem{Blaszczyszyn2015optimal}
B.~Blaszczyszyn and A.~Giovanidis, ``Optimal geographic caching in cellular
  networks,'' in \emph{proc. IEEE ICC}, 2015.

\bibitem{cuiying}
Y.~Cui, D.~Jiang, and Y.~Wu, ``Analysis and optimization of caching and
  multicasting in large-scale cache-enabled wireless networks,'' \emph{IEEE
  Trans. Wireless Commun.}, vol.~15, no.~7, pp. 5101--5112, July 2016.

\bibitem{chenzhen}
Z.~Chen, J.~Lee, T.~Q.~S. Quek, and M.~Kountouris, ``Cooperative caching and
  transmission design in cluster-centric small cell networks,'' \emph{IEEE
  Trans. Wireless Commun.}, vol.~16, no.~5, pp. 3401--3415, May 2017.

\bibitem{xiuhua}
X.~Li, X.~Wang, K.~Li, Z.~Han, and V.~C. Leung, ``Collaborative multi-tier
  caching in heterogeneous networks: Modeling, analysis, and design,''
  \emph{IEEE Trans. Wireless Commun.}, early access, 2017.

\bibitem{TMX}
X.~Xu and M.~Tao, ``Modeling, analysis, and optimization of coded caching in
  small-cell networks,'' \emph{IEEE Trans. Commun.}, vol.~65, no.~8, pp.
  3415--3428, Aug. 2017.

\bibitem{zhangrui}
Y.~Guo, L.~Duan, and R.~Zhang, ``Cooperative local caching under heterogeneous
  file preferences,'' \emph{IEEE Trans. Commun.}, vol.~65, no.~1, pp. 444--457,
  Jan. 2017.

\bibitem{zink2009characteristics}
M.~Zink, K.~Suh, Y.~Gu, and J.~Kurose, ``Characteristics of youtube network
  traffic at a campus network--measurements, models, and implications,''
  \emph{Computer networks}, vol.~53, no.~4, pp. 501--514, 2009.

\bibitem{ahlehagh2014video}
H.~Ahlehagh and S.~Dey, ``Video-aware scheduling and caching in the radio
  access network,'' \emph{IEEE/ACM Trans. Netw.}, vol.~22, no.~5, pp.
  1444--1462, Oct. 2014.

\bibitem{ekstrand2011collaborative}
M.~D. Ekstrand, J.~T. Riedl, J.~A. Konstan \emph{et~al.}, ``Collaborative
  filtering recommender systems,'' \emph{Foundations and
  Trends{\textregistered} in Human--Computer Interaction}, vol.~4, no.~2, pp.
  81--173, 2011.

\bibitem{liujuan}
J.~Liu, B.~Bai, J.~Zhang, and K.~B. Letaief, ``Cache placement in {Fog}-{RAN}s:
  From centralized to distributed algorithms,'' \emph{IEEE Trans. Wireless
  Commun.}, early access, 2017.

\bibitem{traffic}
U.~Paul, A.~P. Subramanian, M.~M. Buddhikot, and S.~R. Das, ``Understanding
  traffic dynamics in cellular data networks,'' in \emph{Proc. IEEE INFOCOM},
  2011.

\bibitem{traffic2}
J.~Yang, Y.~Qiao, X.~Zhang, H.~He, F.~Liu, and G.~Cheng, ``Characterizing user
  behavior in mobile internet,'' \emph{IEEE Trans. Emerg. Topics Comput},
  vol.~3, no.~1, pp. 95--106, Mar. 2015.

\bibitem{CBQ}
B.~Chen and C.~Yang, ``Caching policy optimization for {D2D} communications by
  learning user preference,'' in \emph{Proc. IEEE VTC Spring}, 2017.

\bibitem{bigdata}
E.~Zeydan, E.~Bastug, M.~Bennis, M.~A. Kader, I.~A. Karatepe, A.~S. Er, and
  M.~Debbah, ``Big data caching for networking: moving from cloud to edge,''
  \emph{IEEE Communications Magazine}, vol.~54, no.~9, pp. 36--42, Sept. 2016.

\bibitem{boyd2004convex}
S.~Boyd and L.~Vandenberghe, \emph{Convex optimization}.\hskip 1em plus 0.5em
  minus 0.4em\relax Cambridge university press, 2004.

\bibitem{gamma}
D.~Jaramillo-Ramírez, M.~Kountouris, and E.~Hardouin, ``Coordinated
  multi-point transmission with imperfect {CSI} and other-cell interference,''
  \emph{IEEE Trans. Wireless Commun.}, vol.~14, no.~4, pp. 1882--1896, Apr.
  2015.

\end{thebibliography}

\end{document}